\documentclass[3p]{elsarticle}
\journal{Physics Letters B}

\usepackage{amsmath}
\usepackage{amssymb}
\usepackage{graphicx}
\usepackage{ulem}

\newcommand{\be}{\begin{equation}}
\newcommand{\ee}{\end{equation}}
\newcommand{\ba}{\begin{eqnarray}} 
\newcommand{\ea}{\end{eqnarray}} 
\newcommand{\ban}{\begin{eqnarray*}}
\newcommand{\ean}{\end{eqnarray*}}

\begin{document} 

\title{Two- versus three-body approach \\ to femtoscopic hadron-deuteron correlations}

\author{Stanis\l aw Mr\' owczy\' nski} 

\address{National Centre for Nuclear Research, ul. Pasteura 7,  PL-02-093  Warsaw, Poland}

\date{March 18, 2025}

\begin{abstract}

The three-body approach to hadron-deuteron correlations is shown to turn into a two-body approach if the three-particle hadron-deuteron wave function factorizes into the deuteron wave-function and the wave function of a hadron motion relative to the deuteron. Then, the hadron-deuteron correlation function is as in the two-body approach only the source radius somewhat changes. For this reason, as we argue, the two-body approach works well for kaon-deuteron correlations but it fails for proton-deuteron ones in case of small sources. Applying the three-body approach generalized to the case where the radius of the hadron source is different from the nucleon source radius, we derive the source radius formula which used in the two-body approach gives the correlation function as in the `factorized' three-body one. The formula is discussed in the context of existing and future experimental data. 

\end{abstract}

\maketitle

\section{Introduction}

Femtoscopic correlations of light nuclei produced in heavy-ion collisions have been studied for years at collision energies of tens to hundreds of MeV per nucleon in fixed-target experiments, see the reviews \cite{Boal:1990yh,Verde:2006dh}. In the theoretical description of the correlation functions, light nuclei are treated as point-like objects which is justified for sufficiently large sources of nuclei, larger than a deuteron or alpha particle. The observed correlations are due to the interaction of light nuclei in the reaction final state. 

Measurements of hadron-deuteron correlations in proton-proton collisions at $\sqrt{s} = 13$~TeV \cite{ALICE:2023bny} have caused a revision of the theoretical approach to the femtoscopic correlations. Deuterons produced in these collisions are not fragments of incoming nuclei, but are genuinely produced -- the kinetic energy of colliding protons is converted into the masses of nucleons which form deuterons. Even more significant is the fact that the source of particles in proton-proton collisions is significantly smaller than a deuteron. 

It has been realized \cite{Mrowczynski:2019yrr} that deuterons from  proton-proton collisions cannot be considered as structureless point objects, and that the interaction responsible for the observed correlation occurs simultaneously with the deuteron formation process. The understanding of these two facts has led to the formulation of the three-body approach to hadron-deuteron correlations \cite{Mrowczynski:2019yrr}. Subsequently, the approach has been generalized to $p$-${^3}{\rm He}$ and deuteron-deuteron  correlations where one deals with the four-body problem, see \cite{Bazak:2020wjn,Mrowczynski:2021bzy} and the review \cite{Mrowczynski:2020ugu}.

We emphasize that the necessity of using the three-body approach to the proton-deuteron correlations in proton-proton collisions results not only from the questionable applicability of the two-body approach, but also from the qualitative failure of this approach in describing the experimental correlation function \cite{ALICE:2023bny}. Only sophisticated three-body calculations \cite{Viviani:2023kxw} with the realistic nucleon-nucleon potential and properly antisymmetrized three-nucleon wave function allow one for a correct reproduction of the $p$-$d$ correlation function. 

It should be also noted that, in contrast to the proton-deuteron case, the two-body approach works well in describing the kaon-deuteron correlation function in proton-proton collisions \cite{ALICE:2023bny}. The approach also works for kaon-deuteron, proton-deuteron and deuteron-deuteron correlations in Pb-Pb or Au-Au collisions, see \cite{Rzesa:2024dru} and \cite{STAR:2024lzt}. The approach has been recently refined \cite{Rzesa:2024oqp,Torres-Rincon:2024znb} to better describe the experimental data. However, its applicability remains questionable.

The aim of this paper is twofold. First, we intend to explain why the two-body approach works well for kaon-deuteron correlations in proton-proton collisions, but badly fails for proton-deuteron correlations. The three-body approach turns into the two-body approach if the three-particle hadron-deuteron wave function factorizes into the deuteron wave-function and the wave function of a hadron motion relative to the deuteron \cite{Mrowczynski:2019yrr}. Then, the hadron-deuteron correlation function is as in the two-body approach only the source radius somewhat changes. We argue that the factorization is justified even for small sources for the kaon-deuteron system, but in the case of the proton-deuteron pair the source must be sufficiently large.

Our second goal is to generalize the three-body approach to hadron-deuteron correlations to the case where the radius of the hadron source is different from the nucleon source radius. This problem is particularly important when studying pion-deuteron correlations. Since femtoscopic correlations occur between particles moving with almost the same velocity, the momenta of pions correlated with deuterons are much smaller (due to the large mass difference) than the momenta of nucleons constituting a deuteron. Consequently, the pion source is much larger than that of nucleons, see e.g. \cite{ALICE:2015hvw}.

Although hadron-deuteron correlations have been successfully measured in proton-proton collisions at the LHC \cite{ALICE:2023bny}, measuring deuteron-deuteron correlations is much more difficult. Since deuteron production is a rare event,  production of two deuterons is even rarer. Such data are not available yet and there is no experimental information on single-particle source of deuterons which is required by the two-body approach. In such a case, the hadron-deuteron correlation is described in terms of a hadron-deuteron {\it relative source} which is a convolution of the single-particle hadron and deuteron sources. The radius of the relative source is obtained by fitting a theoretical correlation function to the experimental one. So, a free parameter is in the two-body approach.

In the three-body approach, the hadron-deuteron correlation function is determined by the hadron and nucleon source functions but not of the deuteron one. If the hadron and nucleon source functions are known, and usually they are, the hadron-deuteron correlation function is determined with no free parameter. For the limiting case when the three-body approach changes into the two-body one -- we call it the {\it factorized three-body approach}, we derive the relative source radius of the hadron and deuteron which is expressed through the source radii of the hadron and nucleon. Therefore, we obtain the source function which is needed in the two-body approach. 

The formula of the relative source radius is discussed in the context of experimental data. In particular, we show that 
the kaon-deuteron correlations in proton-proton collisions are described equally well by the two-body approach approach and the factorized three-body one. There are presented predictions of the source radii to be obtained from the $p$-$d$, $d$-$d$ and $p$-${^3}{\rm He}$ correlation functions. 

\section{Two-body approach}
\label{sec-two-body}

To set the stage for our further considerations we first formulate the well-known two-body approach to hadron-deuteron  correlations. The correlation function is defined as
\be
\frac{dP_{hd}}{d^3p_h  d^3p_d} = C({\bf p}_h , {\bf p}_d) \, \frac{dP_h}{d^3p_h}  \frac{dP_d}{d^3p_d} ,
\ee
where $\frac{dP_h}{d^3p_h}$, $\frac{dP_d}{d^3p_d}$ and $\frac{dP_{hd}}{d^3p_h d^3p_d}$ are probability densities to observe $h$, $d$ and $h$-$d$ pairs with momenta ${\bf p}_h$, ${\bf p}_d$ and $({\bf p}_h, {\bf p}_d)$. 

If the correlation is due to final state interactions, the correlation function is, see e.g. \cite{Boal:1990yh,Verde:2006dh},
\be
\label{def-fun-cor}
C({\bf p}_h, {\bf p}_d) = \int d^3 r_h \, d^3 r_d \, 
S_h ({\bf r}_h) \, S_d ({\bf r}_d) |\psi({\bf r}_h,{\bf r}_d)|^2 ,
\ee
where the source function $S_i({\bf r}_i)$ with $i= h, d$ is the normalized probability distribution of emission points and $\psi({\bf r}_h,{\bf r}_h)$ is the wave function of the hadron-deuteron pair in a scattering state. 

To eliminate the center-of-mass motion of the $h$-$d$ pair from the correlation function (\ref{def-fun-cor}), one introduces the center-of-mass variables. Working in the center-of-mass frame it can be done in a non-relativistic manner as femtoscopic correlations occur between particles moving with a small relative velocity. Thus, one writes
\be
\label{CM-2-variables}
\left\{ \begin{array}{ll}
{\bf R} \equiv\frac{m_h{\bf r}_h + m_d{\bf r}_d}{M} ,
\\[1mm]
{\bf r}_{hd} \equiv {\bf r}_h-{\bf r}_d ,
\end{array} \right.
\hspace{1cm}
\left\{ \begin{array}{ll}
{\bf r}_h= {\bf R}+ \frac{m_d}{M}{\bf r}_{hd},
\\[1mm]
{\bf r}_d = {\bf R}-\frac{m_h}{M} {\bf r}_{hd},
\end{array} \right.
\ee
where $M \equiv m_h + m_d$. The wave function is of the form $\psi({\bf r}_h,{\bf r}_d) = e^{i {\bf R}{\bf P}}\phi_{\bf q} ({\bf r}_{hd})$ with ${\bf P} \equiv {\bf p}_h + {\bf p}_d$ being the momentum of the center of mass of the $h$-$d$ system and ${\bf q} \equiv \frac{m_h {\bf p}_d - m_d{\bf p}_h }{M}$ the momentum in the center-of-mass frame. (Actually,  ${\bf P} =0$ in the center-of-mass frame.) The correlation function gets the form of the Koonin-Pratt formula 
\be
\label{Koonin-Pratt}
C({\bf q}) = \int d^3 r \, S^r_{hd} ({\bf r}) |\phi_{\bf q}({\bf r})|^2 ,
\ee
where the {\it relative source} is 
\be
\label{D-r-source-def}
S^r_{hd} ({\bf r}) = \int d^3 R \, S_h \bigg ({\bf R} + \frac{m_d}{M}{\bf r}\bigg) 
\, S_d \bigg ({\bf R} -\frac{m_h}{M} {\bf r} \bigg) .
\ee

Further on we assume that the single-particle source function is Gaussian
\be
\label{D-Gauss}
S_i ({\bf r}) = \bigg({\frac{1}{2\pi R_i^2}}\bigg )^{3/2} e^{ -\frac{{\bf r}^2}{2R_i^2} } ,
\ee
where $\sqrt{3}R_i$ is the root-mean-square (RMS) radius of the single-particle source. 

Using the integral formula
\be
\label{integral}
\int d^3 R \, \exp\big[ - \alpha {\bf R}^2 + \beta {\bf R} \cdot {\bf r} \big] 
= \Big(\frac{\pi}{\alpha}\Big)^{3/2} e^{\frac{\beta^2{\bf  r}^2}{4\alpha}} ,
\ee
where $\alpha, \beta$ are real numbers and $\alpha>0$, the relative source (\ref{D-r-source-def}) is found as 
\be
\label{D-r-source-final}
S^r_{hd} ({\bf r}) = \bigg(\frac{1}{2\pi R_{hd}^2}\bigg )^{3/2} e^{ -\frac{{\bf r}^2}{2R_{hd}^2}} ,
\ee
where
\be
R_{hd} \equiv \sqrt{R_h^2 + R_d^2} ,
\ee
which is independent of particle masses. 

The source radius of a given hadron type is usually obtained from measurements of the $h$-$h$ correlation function. Then, $R_{hh} = \sqrt{2} \,R_h$. When the two-body approach is applied to hadron-deuteron correlations, the deuteron-deuteron correlation function is often not available, and consequently the deuteron source radius $R_d$ is not known. Then, one uses the source function (\ref{D-r-source-final}) and the radius $R_{hd}$ is treated as a free parameter which is obtained by fitting the theoretical correlation function to the experimental one, see e.g. \cite{Rzesa:2024dru}. In the subsequent section we show that the three-body approach is free of this problem.  

\section{Three-body approach}
\label{sec-three-body}

Taking into account that a deuteron is a bound state of neutron and proton created due to final state interactions similarly as the $h$-$d$ correlations, the correlation function is defined as
\be
\frac{dP_{hd}}{d^3p_h \, d^3p_d} = C({\bf p}_h , {\bf p}_d) \, \mathcal{A} \, \frac{dP_h}{d^3p_h} \frac{dP_n}{d^3p_n} \frac{dP_p}{d^3p_p} , ~~~~~~~~{\bf p}_n = {\bf p}_p = \frac{1}{2} \,{\bf p}_d ,
\ee
where except the symbols already introduced there is $\mathcal{A}$ which is the deuteron formation rate defined as 
\be
\frac{dP_d}{d^3p_d} = \mathcal{A} \, \frac{dP_n}{d^3 (p_d / 2)} \frac{dP_p}{d^3(p_d / 2)} .
\ee
The formation rate is known to be \cite{Sato:1981ez} 
\be
\label{D-rate}
\mathcal{A} =\frac{3}{8} (2\pi)^3 \int d^3r_n \, d^3 r_p \, 
S_N({\bf r}_p) \, S_N({\bf r}_n) |\psi_d({\bf r}_n, {\bf r}_p)|^2 ,
\ee
where $\psi_d({\bf r}_n, {\bf r}_p)$ is the deuteron wave function and $S_N({\bf r})$ is the source function of nucleons. The neutrons and protons are assumed to be unpolarized and the spin factor $3/4$ takes into account the fact that there are 3 spin states of a spin-one deuteron and 4 spin states of a nucleon pair. The additional factor 1/2 is included in the formula (\ref{D-rate}) as the neutron–proton pair can be in two isospin states $I = 1,\; I_3 = 0$ and $I = I_3 = 0$ while only the second one contributes to the deuteron production. 

Expressing the deuteron wave function with the center-of-mass variables as
\be
\psi_d({\bf r}_n, {\bf r}_p) = e^{i{\bf P}{\bf R}} \varphi_d ({\bf r}_{np}),
\ee
the deuteron formation rate (\ref{D-rate}) equals
\be
\label{D-rate-relative}
\mathcal{A} 
= \frac{3}{8} (2 \pi)^3 \int d^3r \, S^r_{np} ({\bf r}) |\varphi_d({\bf r})|^2 ,
\ee
where the relative nucleon source $S^r_{np} ({\bf r})$ is defined analogously to Eq.~(\ref{D-r-source-final}). For the Gaussian single-particle source (\ref{D-Gauss}), it is
\be
\label{D-r-Gauss}
S^r_{np} ({\bf r}) = \bigg (\frac{1}{4\pi R_N^2}\bigg )^{3/2} e^{ -\frac{{\bf r}^2}{4R_N^2} } .
\ee

Using essentially the same arguments that lead to formula (\ref{def-fun-cor}), one finds the $h$-$d$ correlation function 
multiplied by the deuteron formation rate ${\cal A}$ in the following form
\be
\label{fun-corr-pi-D-A}
C({\bf p}_h , {\bf p}_d) \, \mathcal{A} 
= \frac{3}{8} (2 \pi )^3 \int d^3 r_h \, d^3 r_n \, d^3 r_p \,
S_N({\bf r}_n) \, S_N({\bf r}_p) \, S_h({\bf r}_h) 
|\psi_{hd}({\bf r}_h, {\bf r}_n, {\bf r}_p)|^2 ,
\ee
where $\psi_{hd}({\bf r}_h, {\bf r}_n, {\bf r}_p)$ is the three-particle wave function of hadron and deuteron. 

To eliminate the center-of-mass motion of the $h$-$d$ pair from the formula (\ref{fun-corr-pi-D-A}), we introduce the Jacobi variables of a three-particle system 
\be 
\label{Jacobi-3}
\left\{ \begin{array}{ll}
{\bf R}=\frac{m_N{\bf r}_n + m_N{\bf r}_p + m_h{\bf r }_h}{M} ,
\\[1mm]
{\bf r}_{np} = {\bf r}_n-{\bf r}_p ,
\\[1mm]
{\bf r}_{hd} = {\bf r}_h - \frac{{\bf r}_n + {\bf r}_p}{2} ,
\end{array} \right.
~~~~~~~~~~~~~~~~~~
\left\{ \begin{array}{ll}
{\bf r}_n = {\bf R} + \frac{1}{2}{\bf r}_{np} - \frac{m_h}{M} {\bf r}_{hd} ,
\\[1mm]
{\bf r}_p = {\bf R} -\frac{1}{2} {\bf r}_{np} - \frac{m_h}{M} {\bf r}_{hd},
\\[1mm]
{\bf r}_h= {\bf R} + \frac{m_d}{M} {\bf r}_{hd} ,
\end{array} \right.
\ee
where the nucleon mass is the same for proton and neutron and $M \equiv 2 m_N +m_h$. Writing down the wave function as
\be
\psi_{hd}({\bf r}_h, {\bf r}_n, {\bf r}_p) = e^{i{\bf P}{\bf R}} \psi_{hd}^{\bf q}({\bf r}_{hd},{\bf r}_{np}),
\ee
where ${\bf q}$ is the hadron momentum in the center-of-mass frame of $h$ and $d$, the correlation function from Eq.~(\ref{fun-corr-pi-D-A}) equals 
\be
\label{fun-corr-pi-D-A-2}
C({\bf q}) = \frac{3}{8} \frac{(2\pi)^3}{\mathcal{A}}
 \int d^3 r_h \, d^3 r_{np}\, S_{hNN}^r({\bf r}_h,{\bf r}_{np}) \,
|\psi_{hd}^{\bf q}({\bf r}_{hd},{\bf r}_{np})|^2 ,
\ee
where
\be
\label{source-hNN}
S_{hNN}^r({\bf r}_h,{\bf r}_{np}) \equiv  
\int d^3 R \, S_N \big({\bf R} + \frac{1}{2} {\bf r}_{np} - \frac{m_h}{M} {\bf r}_{hd}\big) \, 
                   S_N\big({\bf R} - \frac{1}{2} {\bf r}_{np} - \frac{m_h}{M} {\bf r}_{hd}\big) \, 
                   S_h \big({\bf R} + \frac{2m_N}{M} {\bf r}_{hd}\big) .
\ee
The formula (\ref{fun-corr-pi-D-A-2}), where the deuteron formation rate $\mathcal{A}$ is present, is the analog of the  two-body Koonin-Pratt formula (\ref{Koonin-Pratt}). It is the starting point of the full three-body calculations \cite{Viviani:2023kxw} which successfully describe the proton-deuteron correlations in proton-proton collisions at the LHC with no free parameter. 

It is worth noting that the three-body approach to hadron-deuteron correlations is very similar to the approach to the correlations of  three-particles in a scattering state that have recently been measured in case of $p$-$p$-$p$, $p$-$p$-$\bar{p}$, $p$-$p$-$\Lambda$ \cite{ALICE:2022boj}, $p$-$p$-$K$ \cite{ALICE:2023gxp} and $p$-$p$-$\pi$ \cite{ALICE:2025aur} systems. The starting point for calculating the three-particle correlation function is a formula close to Eq.~(\ref{fun-corr-pi-D-A}), which after eliminating the center-of-mass motion becomes analogous to Eq.~(\ref{fun-corr-pi-D-A-2}), except that the wave function does not describe the motion of the hadron relative to the deuteron, but the relative motion of the three particles. The source function (\ref{source-hNN}) can be directly used in the calculation of the $N$-$N$-$h$ correlation function. However, the studies \cite{Kievsky:2023maf,Garrido:2024pwi}, where the $N$-$N$-$N$ and $N$-$N$-$\Lambda$ correlation functions have been calculated, clearly show that the real challenge is not to find the source function, but the three-particle wave function that takes into account the effect of genuine three-particle forces.

\section{From three- to two-body approach}
\label{sec-3-to-2-body}

When the hadron and deuteron are well separated from each other, the three-body approach to hadron-deuteron correlations is expected to change into the two-body approach. Indeed, if the hadron and deuteron are not only well separated but also not  quantum entangled, the wave function of the hadron and deuteron factorizes as 
\be
\label{factorization}
\psi_{hd}^{\bf q}({\bf r}_{hd},{\bf r}_{np}) = \psi_{hd}^{\bf q}({\bf r}_{hd}) \, 
\varphi_d ({\bf r}_{np}) ,
\ee
where $\varphi_d ({\bf r}_{np})$ is the deuteron wave function and $\psi_{hd}^{\bf q}({\bf r}_{hd})$ is that of the hadron and deuteron relative motion. Since the hadron and deuteron are part of a many-particle system, strong decoherence effects are to be expected, and the lack of quantum entanglement is then a natural consequence.

We assume additionally that the single particle source functions of hadron and of nucleon are of the Gaussian from (\ref{D-Gauss}). Using again the integral formula (\ref{integral}), one finds that the source function $S_{hNN}^r({\bf r}_h,{\bf r}_{np})$ factorizes similarly as the wave function in Eq.~(\ref{factorization}) that is
\be
\label{factorization-sources}
S_{hNN}^r({\bf r}_h,{\bf r}_{np}) = S_{np}^r ({\bf r}_{np}) \, S_{hd}^{3r}({\bf r}_{hd}) ,
\ee
where the source function $ S_{np}^r ({\bf r})$ is given by Eq.~(\ref{D-r-Gauss}) and $S_{hd}^{3r}({\bf r})$, which is the relative source of hadron and deuteron in the three-body approach, equals
\be
\label{source-3-hd}
S_{hd}^{3r} ({\bf r}) = \frac{1}{\pi^{3/2} (R_N^2 + 2 R_h^2)^{3/2}} \,
e^{- \frac{{\bf r}^2}{R_N^2 + 2 R_h^2}}. 
\ee
One sees that $S_{hd}^{3r} ({\bf r})$ differs from the analogous source function (\ref{D-r-source-final}) in the two-body approach. We also note that when $R_N = R_h = R_s$ the source function $S_{hd}^{3r}({\bf r})$ equals
\be
\label{source-r-pi-D}
S_{hd}^{3r}({\bf r}) = \Big(\frac{1}{3 \pi R_s^2}\Big)^{3/2} e^{-\frac{{\bf r}^2}{3R_s^2}} ,
\ee
which is the result obtained in \cite{Mrowczynski:2019yrr}.

Substituting the factorization formulas (\ref{factorization}) and (\ref{factorization-sources}) into Eq.~(\ref{fun-corr-pi-D-A-2}), one finds the hadron-deuteron correlation function as
\be
\label{fun-corr-pi-D-bound}
C({\bf q}) = \int d^3 r \, 
S^{3r}_{hd}({\bf r}) \, |\psi_{hd}^{\bf q} ({\bf r})|^2,
\ee
where the deutron formation rate ${\cal A}$ has canceled out. 

If the  factorization relations (\ref{factorization}) and (\ref{factorization-sources}) hold the three-body correlation function (\ref{fun-corr-pi-D-A-2}) is of the same form as the Koonin-Pratt formula (\ref{Koonin-Pratt}) but the source function somewhat differs. When one uses the Koonin-Pratt formula (\ref{Koonin-Pratt}) with the source function (\ref{D-r-source-final}), the two- and three-body formulas (\ref{Koonin-Pratt}) and (\ref{fun-corr-pi-D-bound}) provide exactly the same correlation function if 
\be
\label{hd-radius}
R_{hd} = \sqrt{R_h^2 + \frac{1}{2} R_N^2}. 
\ee
To reliably test the formula (\ref{hd-radius}), the radii $R_{hd}$, $R_h$ and $R_n$ should be measured in collisions at the same collisions energy and centrality class, and the transverse momenta of hadrons $h$ and nucleons $N$ should such be as those in the correlated $h$-$d$ pairs. 

\section{Confrontation with experiment}
\label{sec-discussion}

As already mentioned, while the proton-deuteron correlation function requires the three-body description, the kaon-deuteron correlation function -- both measured in proton-proton collisions at $\sqrt{s} = 13$~TeV -- is well described within the two-body approach \cite{ALICE:2023bny}. This means that the factorization (\ref{factorization}) holds for the $K$-$d$ system but not for the $p$-$d$ one. There is a natural explanation for this fact. The system $p$-$d$ consists of one neutron and two protons with momenta, say, ${\bf p}_1$ and ${\bf p}_2$. When all three particles are localized in a small volume the $p$-$d$ wave function has two terms. In the first one the proton with momentum ${\bf p}_1$  is bound to the neutron, and in the second term the proton with ${\bf p}_2$ is bound to the neutron. Then, the wave function does not factorize according to Eq.~(\ref{factorization}). The factorization is possible for a big source when one proton is significantly closer to the neutron than the other proton. It is also clear that the assumption of factorization is easier to satisfy for the kaon-deuteron than the proton-deutereon system, as there is no ambiguity which particle should be bound to the neutron. 

The kaon-deuteron correlation function obtained in \cite{ALICE:2023bny} has been well described using the Koonin-Pratt formula with the radius $R_{Kd}^{\rm exp} = 1.35^{+0.04}_{-0.05}$~fm. The nucleon source radius has been estimated as $R_N = 1.43 \pm 0.16$~fm \cite{ALICE:2023bny}. The kaon source radius read from the left panel of Fig.~6 of the study \cite{ALICE:2021ovd} is $R_K = 1.0 \pm 0.15$~fm. The average transverse momentum of kaons in the $K$-$d$ measurements \cite{ALICE:2023bny} is about 0.38 GeV while that in $K$-$K$ study \cite{ALICE:2021ovd} is 0.6 GeV. The difference should not significantly influence the value of $R_K$. Substituting $R_K$ and $R_N$ into the formula (\ref{hd-radius}) yields $R_{Kd} = 1.4 \pm 0.2$ which agrees with $R_{Kd}^{\rm exp}$. It shows that the kaon-deuteron correlations in proton-proton collisions are described equally well by both the two- and three-body approach. 

It would be desirable to test the formula (\ref{hd-radius}) in heavy-ion collisions where the source radii change in a sizable range. There are preliminary data on kaon-deuteron correlations in Pb-Pb collisions at $\sqrt{s_{NN}} = 5.02$~TeV \cite{Rzesa:2024dru}. The radius $R_{Kd}$ is obtained for three centrality classes but there is not enough information to check the formula (\ref{hd-radius}). The final data on kaon-deuteron and pion-deuteron correlations are expected soon and hopefully the analysis will be possible. 

When correlations of light nuclei like $p$-$d$, $p$-$^3{\rm He}$, $d$-$d$ are studied,  the three-body approach and the four-body one formulated in \cite{Bazak:2020wjn,Mrowczynski:2021bzy} allows one to express the source radii obtained using the Koonin-Pratt formula through the proton source radius as
\be
\label{relations}
R_{pp} = \sqrt{2} R_p, ~~~~~~~
R_{pd} = \sqrt{\frac{3}{2}} R_p, ~~~~~~~
R_{p{^3{\rm He}}} = \sqrt{\frac{4}{3}} R_p,  ~~~~~~~
R_{dd} = R_p .
\ee
Contrary to naive expectations, the relations (\ref{relations}) show that $R_{pp} > R_{pd} > R_{dd}$. In particular, we have $R_{pd}/R_{dd} = \sqrt{3/2}\approx 1.2$. To compare the source radii $R_p$, $R_{pd}$, $R_{dd}$, $R_{p{^3{\rm He}}}$ to each other, the correlation functions should be measured in the collisions at the same energy and centrality class and the average transverse mass per nucleon of the nuclei under consideration should be also the same. 

The $p$-$d$ and $d$-$d$ correlation functions have been recently measured in Au-Au collisions at $\sqrt{s_{NN}} = 3$~GeV \cite{STAR:2024lzt}. The radius $R_{pd}$ is indeed bigger than $R_{dd}$ but the ratio is about 1.7 not 1.2. The collision energy is rather low and deuterons are fragments of incoming nuclei. Then, not all deuterons are formed due to final state interactions and the relations (\ref{relations}) are of limited applicability. In any case it would be desirable to test the relations systematically. 

\section{Conclusions}
\label{sec-conclusions}

The three-body approach to hadron-deuteron correlations changes into the two-body approach if the three-particle hadron-deuteron wave function factorizes into the deuteron wave-function and the wave function of a hadron motion relative to the deuteron. The assumption of factorization is less restrictive for the kaon-deuteron than the proton-deutereon system, as there is no ambiguity which particle should be bound to the neutron. Presumably for this reason the kaon-deuteron correlations in proton-proton collisions at the LHC are successfully described within the two-body approach while the description of the proton-deuteron correlations requires the three-body approach. 

The factorized three-body approach provides the same correlation function as the Koonin-Pratt formula if the source radius in the later approach is provided by the former one. The source radius formula, which connects the two approaches, holds if deuterons and other light nuclei are formed due final state interactions. The formula can be tested experimentally. 



\begin{thebibliography}{99}
\bibitem{Boal:1990yh}
D.~H.~Boal, C.~K.~Gelbke and B.~K.~Jennings,
Rev. Mod. Phys. \textbf{62} (1990) 553.

\bibitem{Verde:2006dh}
G.~Verde, A.~Chbihi, R.~Ghetti and J.~Helgesson,
Eur. Phys. J. A \textbf{30} (2006) 81.

\bibitem{ALICE:2023bny}
S.~Acharya \textit{et al.} [ALICE Collaboration],
Phys. Rev. X \textbf{14} (2024) 031051.

\bibitem{Mrowczynski:2019yrr} 
St.~Mr\'owczy\'nski and P.~S\l o\'n,
Acta Phys.\ Polon.\ B {\bf 51}  (2020) 1739.

\bibitem{Bazak:2020wjn} 
S.~Bazak and St.~Mr\'owczy\'nski,
Eur. Phys. J. A {\bf 56} (2020) 193.

\bibitem{Mrowczynski:2021bzy}
St.~Mr\'owczy\'nski and P.~S\l{}o\'n,
Phys. Rev. C \textbf{104} (2021) 024909.

\bibitem{Mrowczynski:2020ugu}
St. Mr\' owczy\' nski,
Eur. Phys. J. ST \textbf{229} (2020) 3559.

\bibitem{Viviani:2023kxw}
M.~Viviani, S.~K\"onig, A.~Kievsky, L.~E.~Marcucci, B.~Singh and O.~V\'azquez Doce,
Phys. Rev. C \textbf{108} (2023) 064002.

\bibitem{Rzesa:2024dru}
W.~Rz\c{e}sa and [ALICE Collaboration],
Nuovo Cim. C \textbf{47} (2024) 195.

\bibitem{STAR:2024lzt}
[STAR Collaboration],
[arXiv:2410.03436 [nucl-ex]].

\bibitem{Rzesa:2024oqp}
W.~Rz\c{e}sa, M.~Stefaniak and S.~Pratt,
[arXiv:2410.13983 [nucl-th]].

\bibitem{Torres-Rincon:2024znb}
J.~M.~Torres-Rincon, A.~Ramos and J.~Ruf\'\i{},
[arXiv:2410.23853 [nucl-th]].

\bibitem{ALICE:2015hvw}
J.~Adam \textit{et al.} [ALICE],
Phys. Rev. C \textbf{92} (2015) 054908.

\bibitem{Sato:1981ez} 
H.~Sato and K.~Yazaki,
Phys.\ Lett.\ B {\bf 98} (1981) 153.

\bibitem{ALICE:2022boj}
S.~Acharya \textit{et al.} [ALICE],
Eur. Phys. J. A \textbf{59} (2023) 145.

\bibitem{ALICE:2023gxp}
S.~Acharya \textit{et al.} [ALICE],
Eur. Phys. J. A \textbf{59} (2023) 298.

\bibitem{ALICE:2025aur}
S.~Acharya \textit{et al.} [ALICE],
[arXiv:2502.20200 [nucl-ex]].

\bibitem{Kievsky:2023maf}
A.~Kievsky, E.~Garrido, M.~Viviani, L.~E.~Marcucci, L.~Serksnyte and R.~Del Grande,
Phys. Rev. C \textbf{109} (2024) 034006.

\bibitem{Garrido:2024pwi}
E.~Garrido, A.~Kievsky, M.~Gattobigio, M.~Viviani, L.~E.~Marcucci, R.~Del Grande, L.~Fabbietti and D.~Melnichenko,
Phys. Rev. C \textbf{110} (2024) 054004.


\bibitem{ALICE:2021ovd}
S.~Acharya \textit{et al.} [ALICE],
Phys. Lett. B \textbf{833} (2022), 137335.

\end{thebibliography}
\end{document}